\documentclass[11pt]{scrartcl} 
\usepackage{moreverb}
\usepackage[dvips,colorlinks,bookmarksopen,bookmarksnumbered,citecolor=red,urlcolor=red]{hyperref}
\usepackage{dsfont} 

\usepackage{times}
\usepackage{amsmath}
\usepackage{booktabs}
\usepackage{graphicx}
\usepackage{geometry}
\usepackage{setspace}
\usepackage{multirow}

\newcommand\BibTeX{{\rmfamily B\kern-.05em \textsc{i\kern-.025em b}\kern-.08em
T\kern-.1667em\lower.7ex\hbox{E}\kern-.125emX}}

\begin{document}

\title{Firth's logistic regression with rare events: accurate effect estimates AND predictions?}
\date{May 12, 2016}

\author{Rainer Puhr, Georg Heinze, Mariana Nold, Lara Lusa and Angelika Geroldinger}

\maketitle
\begin{abstract}
Firth-type logistic regression has become a standard approach for the analysis of binary outcomes with small samples. Whereas it reduces the bias in maximum likelihood estimates of coefficients, bias towards $1/2$ is introduced in the predicted probabilities. The stronger the imbalance of the outcome, the more severe is the bias in the predicted probabilities. We propose two simple modifications of Firth-type logistic regression resulting in unbiased predicted probabilities. The first corrects the predicted probabilities by a post-hoc adjustment of the intercept. The other is based on an alternative formulation of Firth-types estimation as an iterative data augmentation procedure. Our suggested modification consists in introducing an indicator variable which distinguishes between original and pseudo observations in the augmented data. In a comprehensive simulation study these approaches are compared to other attempts to improve predictions based on Firth-type penalization and to other published penalization strategies intended for routine use. For instance, we consider a recently suggested compromise between maximum likelihood and Firth-type logistic regression. Simulation results are scrutinized both with regard to prediction and regression coefficients. Finally, the methods considered are illustrated and compared for a study on arterial closure devices in minimally invasive cardiac surgery.       
\end{abstract}


\section{Introduction} \label{introduction}
In logistic regression, Firth-type penalisation \cite{fi1993} has gained increasing popularity as a method to reduce the small-sample bias of maximum likelihood (ML) coefficients. Penalising the likelihood function utilizing Jeffreys invariant prior does not only remove the first-order term in the asymptotic bias expansion of ML estimates but also allows to compute reliable, finite estimates of coefficients in the case of separation, where ML estimation fails, see \cite{hs2002}. Though, reducing the bias in the estimates of coefficients comes at the cost of introducing bias in the predicted probabilities as will be illustrated later: whereas ML gives an average predicted probability equal to the observed proportion of events, Firth-type penalization biases the average predicted probability towards $1/2$. This bias of predictions may be non-negligible if events are very rare or very common. The tempting approach to replace ML estimates by Firth-type penalized estimates whenever the number of observations is critically low, maybe indicated by the occurrence of separation, might then even be detrimental if one is not only interested in effect estimates but also in predicted probabilities. Thus, the present paper has two main objectives. First, we want to clarify how relevant the bias in predicted probabilities based on Firth-type penalization is in practice. We investigate the origin of the bias by simple theoretical considerations and empirically quantify it for realistic situations using simulations. Second, we suggest two simple modifications of Firth-type penalization to overcome the bias of predictions and compare them with alternative methods which were proposed for situations of rare events. \\

The bias in predicted probabilities based on Firth-type penalization already becomes apparent in the simple example of logistic regression with a single binary predictor. Assume that we want to investigate the association between a binary outcome $y$ and some binary risk factor $x$ and observe the following counts: 
\begin{table}[htb]
	\centering
		\begin{tabular}{cccc}
		&&\multicolumn{2}{c}{x}\\
			&&\multicolumn{1}{|c}{0}&1\\
			\cline{2-4}
			\multirow{2}{*}{y}&0& \multicolumn{1}{|c} {95} &4\\
			&1&\multicolumn{1}{|c}{5}&1\\
		\end{tabular}
\end{table}

\noindent For $2\times 2$-tables, predicted probabilities obtained by ML estimation are equal to the proportion of events in the two groups. Thus, we obtain ML predicted probabilities of $5\%$ and $20\%$ for $x=0$ and $x=1$, respectively, corresponding to an overall average predicted probability of $5.71\%$. However, Firth-type penalization results in predicted probabilities of $5.44\%$ and $25\%$, respectively, and an average predicted probability of $6.38\%$, i.e.\ in $11.6\%$ overestimation. This results from the implicit augmentation of cell counts by applying Firth-type penalization to this simple case. Here, Firth-type penalization is equivalent to ML estimation after adding a constant of $0.5$ to each cell, cf.\ \cite{hs2002}. (Note that a $2 \times 2$-table constitutes the simplest case of a saturated model.) The four cell count modifications can be interpreted as four pseudo-observations, each with weight $0.5$, which are added to the original data set. Since the pseudo-data have an event rate of $0.5$, Firth-type penalization leads to overestimation of predicted probabilities in case of rare events. \\
The present paper proposes two simple, generally applicable modifications of Firth-type multivariable logistic regression in order to obtain unbiased average predicted probabilities. First, we consider a simple post-hoc adjustment of the intercept. This \textit{Firth-type logistic regression with intercept-correction (FLIC)} does not alter the bias-corrected effect estimates. By excluding the intercept from the penalization, we do not have to trade accuracy of effect estimates for better predictions. \\
The other approach achieves unbiased predicted probabilities by adjusting for an artificial covariate discriminating between original and pseudo-observations in the iterative weighting procedure mentioned above. In this way, this \textit{Firth-type logistic regression with added covariate (FLAC)} recalibrates the average predicted probability to the proportion of events in the original data. \\
In Section \ref{chap:sim} we perform a comprehensive simulation study comparing the performance of FLIC and FLAC to the performance of other published methods aiming at accurate predictions. In particular, we would like to clarify, whether the proposed modifications improve the accuracy of the predicted probabilities conditional on explanatory variables. Furthermore, in the case of FLAC, we investigate whether the unbiasedness of the predicted probability is paid for by an inflation of bias in effect estimates. As comparator methods we consider a compromise between ML and Firth-type logistic regression recently proposed by Elgmati et al., cf.\ \cite{Elgmati15}, penalization by log-$F(1,1)$ \cite{gm2014} or Cauchy priors \cite{Gelman08}, the ``approximate Bayesian'' and the ``approximate unbiased'' method suggested by King and Zeng, see \cite{Ki01}, and ridge regression. These methods are introduced in Section \ref{methods}. In Section \ref{sec:ACDs}, the performance of these methods is illustrated in a study comparing the use of arterial closure devices to conventional surgical access in minimally invasive cardiac surgery.

\section{Methods} \label{methods}
The logistic regression model $P(y_i=1|x_i)=(1+\exp (-x_i \beta))^{-1}$ with $i=1, \ldots N$ associates a binary outcome $y_i \in \{0,1\}$ to a vector of covariate values $x_i=(1, x_{i1}, \ldots , x_{ip})$ using a $(p+1)$-dimensional vector of regression parameters $\beta=(\beta_0, \beta_1, \ldots , \beta_p)^\prime$. The maximum likelihood (ML) estimate $\hat{\beta}_{\textup{ML}}$ is given by the parameter vector maximizing the log-likelihood function $l (\beta)$ and is usually derived by solving the score equations $\partial l / \partial \beta_r=0$ with $r = 0, \ldots ,p$. For ML estimates, the proportion of observed events is equal to the average predicted probability. This can be seen easily from the explicit form of the ML score functions $\partial l / \partial \beta_r= \sum_i (y_i-\pi_i) x_{ir}$, where $\pi_i=(1+\exp (-x_i \beta))^{-1}$ denotes the predicted probability for the $i$-th observation.\\
Firth has shown that penalizing the likelihood function by the Jeffreys invariant prior removes the first-order term in the asymptotic bias of ML coefficient estimates, cf.\ \cite{fi1993} and \cite{hs2002}. Jeffreys invariant prior is given by $|I(\beta)|^{1/2}=|\mathbf{X}'\mathbf{W}\mathbf{X}|^{1/2}$ with $I(\beta)$ the Fisher information matrix, $\mathbf{X}$ the design matrix and $\mathbf{W}$ the diagonal matrix $\mathrm{diag}(\pi_i(1-\pi_i))$. Estimates of coefficients $\hat{\beta}_{\textup{FL}}$ for Firth-type penalized logistic regression (FL) can be found by solving the corresponding modified score equations
\begin{equation} \label{fse}
\sum_{i=1}^N(y_i-\pi_i+h_i(\frac{1}{2} - \pi_i))x_{ir}=0, \quad r=0, \ldots p,
\end{equation}  
where $h_i$ is the $i$-th diagonal element of the hat matrix $\mathbf{W}^{\frac1 2}\mathbf{X}(\mathbf{X}'\mathbf{W}\mathbf{X})^{-1}\mathbf{X}'\mathbf{W}^{\frac1 2}$, see \cite{hs2002}. Equation (\ref{fse}) for $r=0$ reveals that in general the average predicted probability in Firth-type penalized logistic regression is not equal to the observed proportion of events. Since the determinant of the Fisher information matrix is maximized for $\pi_i=1/2$, it is concluded that Firth-type penalization tends to push the predictions towards $1/2$ compared to ML estimation. Thus, in the situation of rare events, Firth type penalization is prone to overestimate predictions. We can gain more insight into the behaviour of bias of FL predictions by interpreting the modified score equations (\ref{fse}) as score equations for ML estimates for an augmented data set. This data set can be created by complementing each original observation $i$ with two pseudo-observations weighted by $h_i/2$ with same covariate values and with response values $y=0$ and $y=1$, respectively. Thus, FL estimates could be obtained by iteratively applying ML estimation to the augmented data. Given that the trace of the hat matrix is always equal to $p+1$ with $p$ the number of explanatory variables, we see that the augmented data contain $(p+1)/2$ more events than the original one. Consequently, the predicted probability by FL, averaged over the observations in the augmented data, is equal to $(k+ (p+1)/2) /(N+p+1)$ with $k$ the number of events. This gives a very rough approximation of the predicted probability averaged over the original observations, being the closer to the true value the more homogeneous the diagonal elements of the hat matrix are. Unsurprisingly, the relative bias in the average predicted probability is larger for smaller numbers of events $k$ and for larger numbers of parameters $p$ -- exactly in the same situations where the application of Firth-type logistic regression is indicated to reduce small-sample bias.  \\

In the following, we suggest two simple modifications of Firth-type logistic regression which provide average predicted probabilities equal to the observed proportion of events, while preserving the ability to deal with separation: \\

First, we consider altering only the intercept of the Firth-type estimates such that the predicted probabilities become unbiased while keeping all other coefficients constant. In practice, estimates for this \textit{Firth-type logistic regression with intercept-correction (FLIC)} can be derived as follows:
\begin{enumerate}
  \item Determine the Firth-type estimates $\hat{\beta}_{\textup{FL}}$.
  \item Calculate the linear predictors $\hat{\eta}_i=\hat{\beta}_{\textup{FL},1}x_{i1}+\dots+\hat{\beta}_{\textup{FL},p}x_{ip}$, omitting the intercept.
  \item Determine the ML estimate $\hat{\gamma}_0$ of the intercept in the logistic model $P(y_i=1)=(1+\exp(-\gamma_0- \hat{\eta}_i))^{-1}$, containing only a single predictor $\hat{\eta}_i$ with regression coefficient equal to one.
  \item The FLIC estimate $\hat{\beta}_{\textup{FLIC}}$ is then given by the Firth-type estimate $\hat{\beta}_{\textup{FL}}$ with the intercept replaced by $\hat{\gamma}_0$, so	$\hat{\beta}_{\textup{FLIC}}=(\hat{\gamma}_0,\hat{\beta}_{\textup{FL},1},\dots,\hat{\beta}_{\textup{FL},p})$.
\end{enumerate}
By definition, for FLIC the average predicted probability is equal to the proportion of observed events. For the example of the $2\times 2$-table in the introduction, i.e.\ $100$ subjects with $5$ events for $x=0$ and $5$ subjects with one event for $x=1$, FLIC yields predicted probabilities of $4.86 \%$ and $22.82\%$, respectively. Similarly as for Firth-type penalized logistic regression, we suggest to use profile (penalized) likelihood confidence intervals for the coefficients estimated by FLIC except for the intercept, see \cite{hs2002}. Approximate Wald-type confidence intervals for the intercept can be derived from the covariance matrix of the model used to estimate the intercept, which contains the linear predictors as offset. This approximation was evaluated in the simulation study, see Section \ref{chap:sim}. \\

Second, we introduce \textit{Firth-type logistic regression with added covariate (FLAC)}. The basic idea is to discriminate between original and pseudo-observations in the alternative formulation of Firth-type estimation as iterative data augmentation procedure, which was described above. For instance, in the case of $2 \times 2$-tables, where FL amounts to ML estimation of an augmented table with each cell count increased by $0.5$, FLAC estimates are obtained by a stratified analysis of the original $2\times 2$-table and the pseudo data, given by a $2 \times 2$-table with each cell count equal to $0.5$. In the general case, FLAC estimates $\beta_{\textup{FLAC}}$ can be obtained as follows: 
\begin{enumerate}
  \item Apply Firth-type logistic regression and calculate the diagonal elements $h_i$ of the hat matrix.
  \item Construct an augmented data set by stacking (i) the original observations weighted by $1$, (ii) the original observations weighted by $h_i/2$ and (iii) the original observations with opposite response values and weighted by $h_i/2$.
  \item Define an indicator variable $g$ on this augmented data set, being equal to $0$ for (i) and equal to $1$ for (ii) and (iii). 
	\item The FLAC estimates $\hat{\beta}_{\textup{FLAC}}$ are then obtained by ML estimation on the augmented data set adding $g$ as covariate. 
\end{enumerate}
If one does not include the indicator variable $g$ as additional covariate in the last step, ML estimation with augmented data will result in ordinary Firth-type estimates. For the example of the $2\times 2$-table in the introduction, i.e.\ $100$ subjects with $5$ events for $x=0$ and $5$ subjects with one event for $x=1$, FLAC yields predicted probabilities of $5.16 \%$ and $16.83\%$, respectively. For data augmentation in $2 \times 2$-tables the idea of discriminating between original and pseudo observations was already explored in \cite{gr2010}. The average predicted probability by FLAC is equal to the proportion of observed events. Confidence intervals for coefficients estimated by FLAC can be deduced from the ML estimation on the augmented data in the last step. Again, we prefer profile likelihood over Wald confidence intervals. \\

One of the main aims of the simulation study in Section \ref{chap:sim} is to investigate whether these adjustments of the average predicted probability are also reflected in improved predicted probabilities at the subject level. Besides ML and classical Firth-type estimation, we compared the performance of FLIC and FLAC to the performance of the following methods:
\begin{enumerate}
\item[-] A \textit{``weakened'' Firth-type penalization (WF)} is proposed by Elgmati et al., where the likelihood is penalized by $|I(\beta)|^{\tau}$ with weight $\tau$ between $0$ (corresponding to ML estimation) and $1/2$ (corresponding to Firth-type penalization), cf.\ \cite{Elgmati15}. This gives a compromise between accuracy of predictions, where Firth's method is outperformed by ML estimation, and the handling of stability and separation issues, which is a strength of Firth-type penalization. We chose the weight $\tau$ equal to $0.1$ as recommended by Elgmati et al. Similarly as ordinary Firth-type penalization, WF estimates could be obtained by iteratively applying ML estimation to an augmented data set, in this case with an event rate of $(k+ (p+1)/10) /(N+p+1)$.   
	\item[-] \textit{Penalizing by log-$F(1,1)$ priors (LF)} amounts to multiplying the likelihood by $\prod \textup{e}^{\beta_j/2} /(1+ \textup{e}^{\beta_j})$, where the product ranges over $j \in \{1,\ldots, p\}$. This type of penalization is regarded a better choice of a ``default prior'' by Greenland and Mansournia \cite{gm2014} compared to methods such as Firth-type penalization or Cauchy priors. We follow their suggestion to omit the intercept from the penalization, resulting in an average predicted probability equal to the proportion of observed events. Quantitative explanatory variables are advised to be scaled in units that are contextually meaningful, cf.\ \cite{gm2014}. Unlike Jeffreys prior, the log-$F(1,1)$ prior does not depend on the correlation between explanatory variables.  
	\item[-] \textit{Penalization by Cauchy priors (CP)} in logistic regression is suggested as ``default choice for routine applied use'' by Gelman et al., see \cite{Gelman08}. Unlike Greenland and Mansournia, they give an explicit recommendation for data preprocessing, which is also implemented in the corresponding \verb+R+ function \verb+bayesglm+ in the package \verb+arm+. All explanatory variables are shifted to have a mean of $0$. Binary variables are coded to have a range of $1$ and all others are scaled to have a standard deviation of $0.5$. Then, all explanatory variables are penalized by Cauchy priors with center $0$ and scale $2.5$. The intercept is assigned a weaker Cauchy prior, with center $0$ and scale $10$, which implicates that the average predicted probability is in general not equal, but very close to the observed proportion of events.  
		\item[-] The \textit{approximate Bayesian method (AB)} suggested by King and Zeng, cf.\ \cite{Ki01}, takes as starting point small-sample bias corrected logistic regression coefficients, such as coefficients estimated by FL $\hat{\beta}_{\textup{FL}}$. Predicted probabilities by AB $\hat{\pi}_{\textup{AB},i}$ are then obtained by averaging the predicted probabilities by FL over the posterior distribution of the coefficients, 
	\begin{equation} \label{defaB}
	\hat{\pi}_{\textup{AB},i}= \int (1+\exp(- x_i \beta^\ast))^{-1} f(\beta^\ast) d \beta^\ast, 
	\end{equation}
	where $f(\beta^\ast)$ denotes the posterior density of the parameter $\hat{\beta}_{\textup{FL}}$, approximated by $\mathcal{N}(\hat{\beta}_{\textup{FL}}, -(\frac{\partial^2 l}{\partial \beta^2} (\hat{\beta}_{\textup{FL}}))^{-1})$. 
	Instead of deriving $\hat{\pi}_{\textup{AB},i}$ by numerical integration, we make use of the approximation $\hat{\pi}_{\textup{AB},i}= \hat{\pi}_{\textup{FL},i} + C_i$, where the correction factor $C_i$ is given by $(0.5-\hat{\pi}_{\textup{FL},i}) h_i$, cf.\ \cite{Ki01}. In general, equality of average predicted probability and observed event rate does not hold for this method, the bias of the average predicted probability by AB is exactly twice as large as by Firth-type penalization. This follows from the fact that the sum over $C_i$ is equal to $\sum_i \hat{\pi}_{\textup{FL},i}-y_i$ according to the modified score equation for the intercept in FL estimation. King and Zeng are aware that their estimator introduces bias into the predicted probabilities but claim that this is compensated by a reduction of the root mean squared error (RMSE). 
	\item[-] The \textit{approximate unbiased estimator (AU)} also introduced by King and Zeng \cite{Ki01} can be understood as a counterpart to the approximate Bayesian method: using the notation introduced above, predicted probabilities by the approximate unbiased method are now defined as $\hat{\pi}_{\textup{AU},i}= \hat{\pi}_{\textup{FL},i} - C_i$. Similarly as we have seen that the bias in the average predicted probability by AB is twice as large as by FL, one can easily verify that the average predicted probability by AU is equal to the observed proportion of events, i.e.\ unbiased. Nevertheless, King and Zeng consider the AB preferable to the AU ``in the vast majority of applications''. One should be aware that the definition of the AU does not ensure that predicted probabilities fall inside the range of $0$ to $1$. 
	\item[-] In \textit{ridge regression (RR)}, the log-likelihood is penalized by the square of the Euclidean norm of the regression parameters, $\beta_1^2 + \ldots +\beta_p^2$, multiplied by a tuning parameter $\lambda$, see \cite{ch1992}. Since the intercept is omitted from the penalty, the average predicted probability is equal to the proportion of observed events. Following Verweij and Van Houwelingen \cite{vh1994}, in our simulation study the tuning parameter $\lambda$ was chosen by minimizing the penalized version of the Akaike's Information Criterion $AIC=-2l(\hat{\beta}) +2 df_e$, where $df_e= \textup{trace}((\frac{\partial^2 l}{\partial \beta^2} (\hat{\beta}) (\frac{\partial^2 l^\ast}{\partial \beta^2} (\hat{\beta}))^{-1}) )$ with $l^\ast$ the penalized log-likelihood denotes the effective degrees of freedom.
Wald-type confidence intervals were deduced from the penalized variance-covariance matrix with fixed tuning parameter. RR was always performed on scaled explanatory variables with standard deviation equal to one, but results are reported on the original scale. 
\end{enumerate} 
The following types of confidence intervals were chosen for the different methods: Wald-type confidence intervals were used for ML, CP and RR and profile likelihood confidence intervals for WF, FL, FLAC and LF. Confidence intervals for FLIC were constructed as explained above.

\section{Simulation study}\label{chap:sim}
The empirical performance of the methods introduced in Section \ref{methods} was evaluated in a comprehensive simulation study. Integral parts were the comparison of bias and RMSE of predicted probabilities and estimates of coefficients. Furthermore, we assessed discrimination, quantified by the c-statistic, and the calibration slope of the models \cite{sb2010}. Confidence intervals were evaluated with regard to power, length and coverage. 
Results on predictions are presented for all methods introduced above, ML, WF, FL, FLIC, FLAC, LF, CP, AU, AB and RR, whereas results on coefficient estimates only concern ML, WF, FL, FLIC, FLAC, LF, CP and RR. Whenever we are solely interested in coefficient estimates except for the intercept, results from FL and FLIC agree and are presented jointly. 

\subsection{Data generation}\label{ch:simdat}
Binary outcomes $y_i$ were generated from the logistic model $P(y_i|x_{i1},\ldots \linebreak[0],x_{i10})=(1+\exp(-\beta_0-\beta_1x_{i1}- \ldots\linebreak[0]- \nolinebreak \beta_{10}x_{i10}))^{-1}, \, i=1 \ldots \linebreak[0] N,$ with ten explanatory variables $x_{i1},\dots,x_{i10}$. The joint distribution of the explanatory variables follows Binder, Sauerbrei und Royston \cite{bs2011} and includes continuous as well as categorical variables. Since the focus of our paper is on predictions, we have reduced the number of explanatory variables from seventeen \cite{bs2011} to ten and considered only linear effects.

First, we generated ten standard normal random variables $z_{ij} \sim \mathcal{N}\left(0,1\right), \; j=1,\ldots \linebreak[0], 10, \; i=1, \ldots \linebreak[0] N $, with correlation structure as listed in Table \ref{tbl:vars} in the Appendix. By applying the transformations described in Table \ref{tbl:vars} to the variables $z_{ij}$, four continuous, four binary and two ordinal variables $x_{ij}$ were derived. The continuous variables $x_{i1},x_{i4},x_{i5}$ and $x_{i8}$ were truncated at the third quartile plus five times the interquartile distance in each simulated dataset.

Coefficients $\beta_2,\; \beta_6,\; \beta_9$ and $\beta_{10}$ of the binary variables were set to $0.69$, coefficients $\beta_3$ and $\beta_7$ of ordinal variables to $0.345$. This corresponds to an odds ratio of $2$ for binary variables and to an odds ratio of $\sqrt{2}$ for consecutive categories of ordinal variables. Coefficients $\beta_1,\;\beta_4,\;\beta_5$ and $\beta_8$ of the continuous variables were chosen such that the difference between the first and the sixth sextile of the empirical distribution function corresponds to an odds ratio of $2$. Finally, the intercept $\beta_0$ was chosen such that a certain proportion of events was obtained. We considered all combinations of sample sizes $N \in \{500, 1400, 3000\}$ and population event rates $\pi \in \{1\%,2\%,5\%,10\%\}$ such that the expected number of events was always greater than $20$. In addition to the covariate effects described above, we considered effects all equal to zero and effects half the size of the coefficients stated above, referred to as an effect size $a$ of $0$, $0.5$ and $1$. Finally, for each of these scenarios we also took into account effects of mixed signs, multiplying the coefficients $\beta_j, j=6,\dots,10$ by $-1$. For each of these $45$ scenarios, $1000$ datasets were simulated and analysed by the methods described in Section \ref{methods}.
The sample sizes have been chosen such that in approximately $50\%$, $80\%$ and $95\%$ of $1000$ datasets with event rate of $10\%$ and large, all positive effects the Firth-type coefficients were significantly different from zero, respectively.

For random variable generation, the function \verb+rmvnorm+ from package \verb+mvtnorm+ \cite{mvtnorm} was used in the case of multivariate normal distributions and the function \verb+rbinom+ from package \verb+base+ in the case of binomial distributions (\verb+R 3.0.3+ \cite{R}). ML and FL were estimated using the \verb+R+-package \verb+logistf+ \cite{logistf}. This package was also used in the implementation of the algorithms for FLIC and FLAC as described in Section \ref{methods}. For WF a self-modified version of \verb+logistf+ was used. RR was estimated by the \verb+R+-package \verb+rms+ \cite{rms}, CP by the function \verb+bayesglm+ in the package \verb+arm+ \cite{arm} and LF was implemented as maximum likelihood analysis of the appropriately modified data set, see \cite{gm2014}, page 3138.

\subsection{Results}
For clarity, tables of results in this section are restricted to some selected scenarios with coefficients of mixed sign, i.e.\ $ \textup{sgn} (\beta_j)=-1, j=6,\dots,10$. Results for scenarios with only positive coefficients $\beta$ were similar and are available from the authors upon request. In the text we refer to all scenarios if not stated otherwise. 
Separation was encountered at most in $4$ out of $1000$ simulated datasets per scenario. Although some methods can handle separation in data sets, we excluded these cases from analyses to retain comparability.\\

We begin by analysing the discrepancy between the estimated event rate, i.e.\ the average predicted probability, and the observed event rate for the methods WF, FL, CP and AB, see Table \ref{tbl:rbias}. All other methods (ML, FLIC, FLAC, LF, AU and RR) give average predicted probabilities equal to the observed event rate by construction. Among all considered scenarios, the scenario with a sample size of $500$, an expected event rate of $0.05$ and with all explanatory variables being unrelated to the outcome (effect size of zero) was associated with the largest relative bias of the event rate for WF $(4\%)$, FL $(19.4\%)$ and for AB $(38.8\%)$. The bias of the estimated event rate for FL was about five times larger than for WF, and exactly half the size of the bias for AB. For CP, the difference between estimated and observed event rate can be considered negligible (relative bias smaller than $0.3\%$ for all scenarios). For all methods, the RMSE was only slightly larger than the bias, i.e.\ the variance of the estimated event rates was small. These numbers show that in cases of rare events the systematic error in the estimated proportion of events can not be ignored.

\begin{table}[htb]
		\caption{Bias and RMSE $(\times 100)$ of the estimated event rate $\overline{\hat{\pi}_i}$ relative to the observed event rate $\overline{y_i}$, for scenarios with small effect size $(a=0.5)$ and with coefficients of mixed sign. The methods not considered in the table (ML, FLIC, FLAC, LF, AU and RR) have zero bias and RMSE by construction.}\label{tbl:rbias}		
	\footnotesize
		\centering
		\begin{tabular}{ccccccccccc}
			\toprule
			\multicolumn{1}{c}{\textbf{N}} & \multicolumn{1}{c}{\textbf{Method}} & \multicolumn{4}{c}{\textbf{Relative bias} } && \multicolumn{4}{c}{\textbf{Relative RMSE}}\\
			 && \multicolumn{4}{c}{$\mathbf{\pi}$} && \multicolumn{4}{c}{$\mathbf{\pi}$}\\
			\cline{3-6} \cline{8-11}
			&  & 0.01 & 0.02 & 0.05 & 0.1 && 0.01 & 0.02 & 0.05 & 0.1 \\ 
			\hline
500 & WF &  &  & 3.7 & 1.6 &  &  &  & 3.8 & 1.6 \\ 
   & FL &  &  & 18.2 & 7.8 &  &  &  & 18.7 & 7.9 \\ 
   & CP &  &  & 0.2 & 0.1 &  &  &  & 0.2 & 0.1 \\ 
   & AB &  &  & 36.4 & 15.5 &  &  &  & 37.4 & 15.8 \\ 
  1400 & WF &  & 3.7 & 1.3 & 0.6 &  &  & 3.8 & 1.3 & 0.6 \\ 
   & FL &  & 18.5 & 6.6 & 2.8 &  &  & 19.0 & 6.7 & 2.8 \\ 
   & CP &  & 0.2 & 0.1 & 0.0 &  &  & 0.3 & 0.1 & 0.0 \\ 
   & AB &  & 37.1 & 13.2 & 5.6 &  &  & 38.0 & 13.3 & 5.7 \\ 
  3000 & WF & 3.6 & 1.7 & 0.6 & 0.3 &  & 3.7 & 1.7 & 0.6 & 0.3 \\ 
   & FL & 17.9 & 8.6 & 3.1 & 1.3 &  & 18.3 & 8.6 & 3.1 & 1.3 \\ 
   & CP & 0.3 & 0.1 & 0.0 & 0.0 &  & 0.3 & 0.1 & 0.0 & 0.0 \\ 
   & AB & 35.9 & 17.1 & 6.2 & 2.6 &  & 36.6 & 17.3 & 6.2 & 2.6 \\ 
			\bottomrule
		\end{tabular}
\end{table}

\begin{figure}[htbp]
	\centering
		\includegraphics{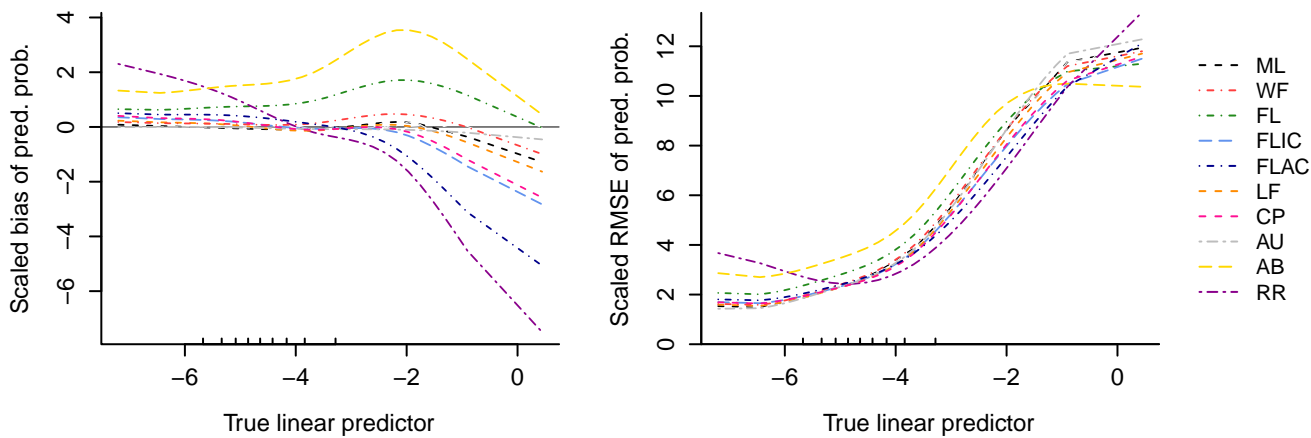}
	\caption{Bias and RMSE of predicted probabilities (scaled by the standard error of proportions $\sqrt{p(1-p)/N}$ with $p$ the probability corresponding to the true linear predictor) by true linear predictor, exemplarily for the scenario $N=1400$, $\bar{y}=0.02$, large effect size ($a=1$) and coefficients of mixed signs. For the calculation of bias and RMSE, the predicted probabilities were splitted into $30$ groups using adequate quantiles of the true linear predictor. Cubic smoothing splines were then fitted to the derived bias and RMSE values in the $30$ groups. Upward directed ticks on the x-axis mark the deciles of the true linear predictor. See Supplementary figure 2 for an unscaled version of this plot. } \label{fig:probbylin}
\end{figure}

Table \ref{tbl:probstats} shows the bias and RMSE of the individual predicted probabilities. ML, FLIC, FLAC, LF, AU and RR give unbiased predicted probabilities -- the fact that the respective numbers are not exactly equal to zero is due to sampling variability in the simulation. RR was associated with considerably lower RMSE than other methods in all but one scenarios, with an RMSE up to $52.4\%$ lower than the second best performing method which was either FLAC or, in some situations with large effect size, FLIC. Both FLAC and FLIC did not only correct the bias in predicted probabilities but also reduced the variance and in particular the RMSE (by up to $23.4\%$ and $16.2\%$, respectively) in comparison to FL. CP resulted in smaller RMSE than LF, which still performed better than WF in all scenarios. The approximate Bayesian method AB performed worst with respect to bias and RMSE in almost all settings. For 17 simulation scenarios with smaller number of events, the approximate unbiased method AU yielded predicted probabilities outside of $[0,1]$, with a minimum value of $-0.02$ and a maximum value of $1.0008$. With increasing sample size, event rate and effect size, differences between methods diminished. \\
All methods except for RR yielded mean calibration slopes smaller than $1$ throughout all scenarios, indicating underestimation of small event probabilities and overestimation of larger ones (cf.\ Table \ref{tbl:probstats}). The method achieving the calibration slope closest to the optimal value of $1$ was either FLAC or RR depending on the scenario. Both FLIC and FLAC outperformed ordinary Firth-type estimation with respect to calibration. With increasing sample size and expected event rate both, differences between methods and the distance to the optimal value of $1$, decreased.\\
Figure \ref{fig:probbylin} investigates the bias and RMSE of predictions in relation to the size of the true linear predictor exemplarily for one simulation scenario. In line with the results on the calibration slope in Table \ref{tbl:probstats}, we find that RR strongly overestimates small event probabilities and underestimates large ones. A similar, but less pronounced pattern is shown by FLAC predictions. Predictions by AU were close to unbiased over the whole range of predictions, but were associated with considerable variance, in particular for larger predictions. For better discriminability of methods, Figure \ref{fig:probbylin} shows the bias and RMSE scaled by the standard error of proportions corresponding to the respective true linear predictors. Supplementary figure 2 is the unscaled version of Figure \ref{fig:probbylin}.

\begin{table}[phtb]
\caption{Mean bias and RMSE $(\times 10000)$ of predicted probabilities $\hat{\pi}_i$, mean and standard deviation ($\times 100)$ of calibration slopes, for selected simulation scenarios with coefficients of mixed signs. (See Supplementary table 1 for further scenarios).} \label{tbl:probstats}
	\footnotesize
	\centering
	\begin{tabular}{rrrp{0.1cm}rrrp{0.05cm}rrrp{0.1cm}rrp{0.05cm}rr}
	\toprule
	&&&& \multicolumn{7}{c}{\textbf{Predictions}} && \multicolumn{5}{c}{\textbf{Calibration slope}}\\
	\multicolumn{1}{c}{\textbf{N}} & \multicolumn{1}{c}{$\mathbf{\pi}$} & \textbf{Method} &&  \multicolumn{3}{c}{\textbf{Bias} ($\times10000$)} &&   \multicolumn{3}{c}{\textbf{RMSE} ($\times10000$)}  && \multicolumn{2}{c} {\textbf{Mean} ($\times 100$)} && \multicolumn{2}{c} {\textbf{SD} ($\times 100$)}\vspace{0.1cm}\\
	& & && \multicolumn{3}{c}{$a$} && \multicolumn{3}{c}{$a$} && \multicolumn{2}{c}{$a$} && \multicolumn{2}{c}{$a$} \\ 
		\cline{5-7} \cline{9-11} \cline{13-14} \cline{16-17}
	& & && 0 & 0.5 & 1 && 0 & 0.5 & 1 &&  0.5 & 1 && 0.5 & 1\\ 
	\midrule
  500 & 0.05 & ML &  & -1 & 0 & -1 &  & 351 & 403 & 469 &  & 43 & 80 &  & 16 & 18 \\ 
   &  & WF &  & 18 & 18 & 14 &  & 359 & 408 & 469 &  & 43 & 80 &  & 16 & 17 \\ 
   &  & FL &  & 91 & 87 & 74 &  & 392 & 430 & 472 &  & 41 & 78 &  & 14 & 16 \\ 
   &  & FLIC &  & -1 & 0 & -1 &  & 332 & 375 & 437 &  & 48 & 87 &  & 17 & 20 \\ 
   &  & FLAC &  & -1 & 0 & -1 &  & 312 & 360 & 435 &  & 50 & 91 &  & 19 & 22 \\ 
   &  & LF &  & -1 & 0 & -1 &  & 340 & 391 & 453 &  & 45 & 83 &  & 17 & 19 \\ 
   &  & CP &  & 0 & 1 & 0 &  & 326 & 377 & 440 &  & 47 & 86 &  & 18 & 20 \\ 
   &  & AU &  & -1 & 0 & -1 &  & 351 & 407 & 473 &  & 43 & 80 &  & 16 & 18 \\ 
   &  & AB &  & 184 & 174 & 150 &  & 457 & 477 & 495 &  & 39 & 76 &  & 12 & 14 \\ 
   &  & RR &  & -1 & 0 & -1 &  & 153 & 282 & 424 &  & 128 & 117 &  & 85 & 66 \\ 
	\cline{2-17}
   & 0.10 & ML &  & -1 & -4 & -2 &  & 463 & 503 & 533 &  & 61 & 87 &  & 16 & 13 \\ 
   &  & WF &  & 16 & 11 & 11 &  & 466 & 504 & 531 &  & 60 & 87 &  & 15 & 13 \\ 
   &  & FL &  & 82 & 71 & 63 &  & 481 & 509 & 529 &  & 60 & 88 &  & 15 & 13 \\ 
   &  & FLIC &  & -1 & -4 & -2 &  & 447 & 481 & 512 &  & 64 & 91 &  & 16 & 14 \\ 
   &  & FLAC &  & -1 & -4 & -2 &  & 434 & 476 & 512 &  & 65 & 93 &  & 17 & 14 \\ 
   &  & LF &  & -1 & -4 & -2 &  & 456 & 495 & 523 &  & 62 & 88 &  & 16 & 13 \\ 
   &  & CP &  & 0 & -3 & -1 &  & 446 & 486 & 514 &  & 63 & 90 &  & 16 & 14 \\ 
   &  & AU &  & -1 & -4 & -2 &  & 463 & 506 & 535 &  & 60 & 87 &  & 16 & 13 \\ 
   &  & AB &  & 164 & 147 & 127 &  & 516 & 526 & 536 &  & 59 & 88 &  & 14 & 12 \\ 
   &  & RR &  & -1 & -4 & -2 &  & 235 & 406 & 506 &  & 116 & 102 &  & 53 & 23 \\ 
	\cline{1-17}
	\cline{1-17}
  3000 & 0.01 & ML &  & 0 & 0 & 0 &  & 66 & 84 & 137 &  & 51 & 85 &  & 17 & 20 \\ 
   &  & WF &  & 4 & 4 & 4 &  & 68 & 86 & 138 &  & 49 & 83 &  & 17 & 19 \\ 
   &  & FL &  & 18 & 18 & 16 &  & 78 & 97 & 144 &  & 45 & 78 &  & 14 & 16 \\ 
   &  & FLIC &  & 0 & 0 & 0 &  & 65 & 82 & 130 &  & 52 & 88 &  & 17 & 21 \\ 
   &  & FLAC &  & 0 & 0 & 0 &  & 60 & 75 & 127 &  & 58 & 97 &  & 20 & 25 \\ 
   &  & LF &  & 0 & 0 & 0 &  & 65 & 82 & 134 &  & 52 & 86 &  & 18 & 21 \\ 
   &  & CP &  & 0 & 0 & 1 &  & 62 & 79 & 130 &  & 54 & 89 &  & 19 & 22 \\ 
   &  & AU &  & 0 & 0 & 0 &  & 66 & 85 & 139 &  & 50 & 84 &  & 17 & 20 \\ 
   &  & AB &  & 36 & 35 & 32 &  & 94 & 114 & 156 &  & 40 & 73 &  & 12 & 14 \\ 
   &  & RR &  & 0 & 0 & 0 &  & 29 & 60 & 125 &  & 135 & 111 &  & 81 & 40 \\ 
	\cline{2-17}
   & 0.02 & ML &  & 0 & 0 & 0 &  & 89 & 110 & 158 &  & 71 & 92 &  & 19 & 14 \\ 
   &  & WF &  & 4 & 4 & 3 &  & 90 & 112 & 158 &  & 70 & 91 &  & 18 & 14 \\ 
   &  & FL &  & 18 & 17 & 15 &  & 96 & 118 & 161 &  & 66 & 89 &  & 16 & 13 \\ 
   &  & FLIC &  & 0 & 0 & 0 &  & 88 & 109 & 154 &  & 71 & 94 &  & 18 & 15 \\ 
   &  & FLAC &  & 0 & 0 & 0 &  & 84 & 104 & 152 &  & 76 & 98 &  & 20 & 16 \\ 
   &  & LF &  & 0 & 0 & 0 &  & 88 & 109 & 156 &  & 72 & 93 &  & 19 & 15 \\ 
   &  & CP &  & 0 & 0 & 0 &  & 86 & 107 & 154 &  & 73 & 94 &  & 19 & 15 \\ 
   &  & AU &  & 0 & 0 & 0 &  & 88 & 111 & 159 &  & 71 & 91 &  & 19 & 14 \\ 
   &  & AB &  & 35 & 34 & 30 &  & 107 & 128 & 166 &  & 62 & 87 &  & 15 & 12 \\ 
   &  & RR &  & 0 & 0 & 0 &  & 48 & 93 & 152 &  & 122 & 102 &  & 53 & 20 \\ 
	\cline{2-17}
   & 0.05 & ML &  & -1 & 0 & 0 &  & 135 & 157 & 189 &  & 86 & 97 &  & 14 & 9 \\ 
   &  & WF &  & 3 & 3 & 3 &  & 135 & 158 & 189 &  & 85 & 96 &  & 14 & 9 \\ 
   &  & FL &  & 16 & 15 & 13 &  & 138 & 160 & 189 &  & 84 & 96 &  & 14 & 9 \\ 
   &  & FLIC &  & -1 & 0 & 0 &  & 134 & 156 & 187 &  & 86 & 98 &  & 14 & 9 \\ 
   &  & FLAC &  & -1 & 0 & 0 &  & 132 & 154 & 187 &  & 88 & 99 &  & 15 & 9 \\ 
   &  & LF &  & -1 & 0 & 0 &  & 134 & 157 & 188 &  & 86 & 97 &  & 15 & 9 \\ 
   &  & CP &  & 0 & 0 & 0 &  & 133 & 156 & 187 &  & 87 & 97 &  & 15 & 9 \\ 
   &  & AU &  & -1 & 0 & 0 &  & 135 & 158 & 189 &  & 86 & 96 &  & 14 & 9 \\ 
   &  & AB &  & 32 & 31 & 26 &  & 145 & 165 & 191 &  & 82 & 96 &  & 13 & 9 \\ 
   &  & RR &  & -1 & 0 & 0 &  & 94 & 149 & 186 &  & 105 & 100 &  & 24 & 10 \\ 
   \bottomrule
	\end{tabular}
\end{table}

The discriminative power of the models (in terms of c-indices) was evaluated with regard to a new, independently generated outcome drawn from the logistic model with the same covariate values as in the training data. Highest discrimination was achieved by either RR or AB depending on the scenario but there was little variation across methods, cf. Supplementary table 2. By construction, FL and FLIC give the same c-indices since adjusting the intercept does not change the order of predicted probabilities. The ``optimal value'' in the Supplementary table 2 was obtained by calculating the c-index based on event probabilities from the true model.

Since the relation between predicted probabilities and linear predictors is non-linear of non-vanishing curvature for linear predictors smaller than $0$, a reduced bias of predictions often comes at the cost of an increased bias of linear predictors and vice versa. This effect is less pronounced if the variability of the estimator is small. For instance, RR, having the smallest RMSE of linear predictors among the investigated methods, performed well with regard to both, linear predictors and predicted probabilities, cf.\ Supplementary table 3. Bias and RMSE of linear predictors were of largest absolute size for ML across all $45$ scenarios. FL was least biased in almost all scenarios but with considerably larger RMSE than RR. 
Again, with increasing number of expected events and effect size differences between methods decreased.

Table \ref{tbl:coefstats} shows the absolute bias and RMSE of the standardized coefficients, averaged over all explanatory variables, omitting the intercept. For simulation scenarios with non-zero covariate effects, FL outperformed the other methods in $28$ out of $36$ scenarios with respect to absolute bias. Even in unfavourable scenarios, its average standardized bias did not exceed $1\%$. Concerning the RMSE, FL ranked at least fourth place, almost always clearly outmatched by RR and throughout all scenarios slightly worse than CP and, interestingly, FLAC, but never worse than LF. ML and WF were associated with the largest and second largest RMSE throughout all simulation scenarios. Unsurprisingly, RR gave the smallest absolute bias in simulation scenarios with zero covariate effects but closely followed by FL.

\begin{table}[htbp]
\caption{Absolute bias and RMSE ($\times1000$) of standardized coefficients, averaged over all explanatory variables (omitting the intercept), for some simulation scenarios with coefficients of mixed signs. (See Supplementary table 4 for further scenarios.)} \label{tbl:coefstats}
	\footnotesize
	\centering
	\begin{tabular}{rrrp{0.1cm}rrrp{0.1cm}rrr}
	\toprule
	\multicolumn{1}{c}{\textbf{N}} & \multicolumn{1}{c}{$\mathbf{\pi}$} & \textbf{Method} &&  \multicolumn{3}{c}{\textbf{Bias} ($\times1000$)} &&    \multicolumn{3}{c}{\textbf{RMSE} ($\times1000$)}\\
	& & && \multicolumn{3}{c}{$a$} && \multicolumn{3}{c}{$a$} \\ 
	\cline{5-7} \cline{9-11}
	& & && 0 & 0.5 & 1 && 0 & 0.5 & 1 \\ 
  \midrule
 500 & 0.05 & ML &  & 23 & 17 & 29 &  & 277 & 266 & 288 \\ 
   &  & WF &  & 19 & 14 & 21 &  & 272 & 261 & 281 \\ 
   &  & FL/FLIC &  & 7 & 5 & 9 &  & 253 & 244 & 259 \\ 
   &  & FLAC &  & 17 & 16 & 16 &  & 239 & 235 & 252 \\ 
   &  & LF &  & 22 & 10 & 12 &  & 265 & 252 & 266 \\ 
   &  & CP &  & 18 & 14 & 24 &  & 245 & 238 & 251 \\ 
   &  & RR &  & 3 & 109 & 124 &  & 78 & 166 & 244 \\ 
	\cline{2-11}
   & 0.10 & ML &  & 11 & 7 & 21 &  & 191 & 188 & 203 \\ 
   &  & WF &  & 10 & 6 & 17 &  & 189 & 186 & 201 \\ 
   &  & FL/FLIC &  & 4 & 3 & 3 &  & 181 & 178 & 191 \\ 
   &  & FLAC &  & 10 & 9 & 7 &  & 177 & 175 & 189 \\ 
   &  & LF &  & 11 & 5 & 9 &  & 187 & 183 & 196 \\ 
   &  & CP &  & 10 & 8 & 10 &  & 179 & 177 & 189 \\ 
   &  & RR &  & 2 & 92 & 75 &  & 68 & 143 & 190 \\ 
	\cline{1-11}
  3000 & 0.01 & ML &  & 22 & 15 & 18 &  & 234 & 223 & 231 \\ 
   &  & WF &  & 19 & 12 & 14 &  & 232 & 221 & 228 \\ 
   &  & FL/FLIC &  & 8 & 6 & 8 &  & 223 & 213 & 218 \\ 
   &  & FLAC &  & 18 & 20 & 17 &  & 208 & 203 & 212 \\ 
   &  & LF &  & 21 & 12 & 8 &  & 227 & 214 & 219 \\ 
   &  & CP &  & 19 & 16 & 18 &  & 214 & 206 & 212 \\ 
   &  & RR &  & 3 & 104 & 102 &  & 71 & 154 & 210 \\ 
	\cline{2-11}
   & 0.02 & ML &  & 10 & 7 & 10 &  & 160 & 152 & 158 \\ 
   &  & WF &  & 9 & 7 & 8 &  & 159 & 151 & 157 \\ 
   &  & FL/FLIC &  & 4 & 5 & 4 &  & 156 & 149 & 153 \\ 
   &  & FLAC &  & 9 & 10 & 7 &  & 151 & 145 & 151 \\ 
   &  & LF &  & 10 & 7 & 4 &  & 158 & 149 & 154 \\ 
   &  & CP &  & 10 & 8 & 9 &  & 153 & 146 & 151 \\ 
   &  & RR &  & 3 & 81 & 60 &  & 64 & 129 & 155 \\ 
	\cline{2-11}
   & 0.05 & ML &  & 5 & 3 & 5 &  & 101 & 98 & 103 \\ 
   &  & WF &  & 4 & 2 & 4 &  & 101 & 97 & 103 \\ 
   &  & FL/FLIC &  & 3 & 2 & 2 &  & 100 & 97 & 102 \\ 
   &  & FLAC &  & 5 & 4 & 4 &  & 98 & 96 & 101 \\ 
   &  & LF &  & 5 & 3 & 3 &  & 100 & 97 & 102 \\ 
   &  & CP &  & 5 & 3 & 4 &  & 99 & 96 & 101 \\ 
   &  & RR &  & 2 & 47 & 25 &  & 60 & 95 & 102 \\ 
	\bottomrule
	\end{tabular}
\end{table}

Finally, the approximate Wald-type standard error for the intercept in FLIC estimation suggested in Section \ref{methods} was evaluated by comparing the corresponding confidence intervals to jackknife and bootstrap ($200$ repetitions) confidence intervals. Due to the computational burden, this comparison was restricted to the $25$ simulation scenarios with sample size $N=500$ and $N=1400$. In $17$ out of $20$ simulation scenarios with non-zero covariate effects the approximate approach yielded the confidence bounds that most often excluded zero
, but differences between methods were marginal, cf.\ Supplementary table 5. The approximate Wald-type confidence interval was also the shortest. Especially in extreme situations, the coverage exceeded the nominal significance level for all three methods. 

For all methods, results on $95\%$ confidence intervals averaged over all coefficients omitting the intercept, can be found in the Supplementary table 6. Coverage was reasonable for all methods except for RR, ranging between $93.9\%$ and $96\%$ across methods and scenarios. RR confidence intervals were overly conservative in simulation scenarios without any covariate effects with coverage levels as high as $99.7\%$ and were too optimistic for scenarios with moderate or large effects. They were clearly shorter with less power to exclude $0$ compared to the other methods. In $29$ out of $36$ simulation scenarios CP and FLAC were among the four methods with smallest power, often combined with a slight conservatism. Though, in general, there were little differences in the behaviour of the confidence intervals among all methods except for RR.

\section{Example: arterial closure devices in minimally invasive cardiac surgery} \label{sec:ACDs}
In a retrospective study at the Department of Cardiothoracic Surgery of the University Hospital of the Friedrich-Schiller University Jena, the use of arterial closure devices (ACDs) in minimally invasive cardiac surgery was compared to conventional surgical access with regard to the occurrence of cannulation-site complications. Of the $440$ patients eligible for analysis, $16$ $(3.6\%)$ encountered complications. About one fifth of surgeries ($90$ cases) were performed with conventional surgical access to the groin vessels. The complication rate was $8.9\%$ ($8$ cases) for the conventional surgical access and $2.3\%$ ($8$ cases) for the ACDs group. For the purpose of illustration, the analysis in the present paper was restricted to four adjustment variables selected by medical criteria, the \textit{logistic EuroSCORE}, which estimates the risk of mortality in percent, the presence of \textit{previous cardiac operations} (yes/no), the body mass index (\textit{BMI}), for which five missing values were replaced by the mean BMI, and the presence of \textit{diabetes} (yes/no). The aim of the study was twofold, first, to estimate the adjusted effect of the surgical access procedure on the occurrence of complications and, second, to quantify the risk for complications. Multivariable logistic regression models with the five explanatory variables \textit{type of surgical access}, \textit{logistic EuroSCORE}, \textit{previous cardiac operations}, \textit{BMI} and \textit{diabetes} were estimated by ML, WF, FL, FLAC, LF, CP and RR. In addition, predicted probabilities were obtained by FLIC, AB and AU. All methods gave significant, large effect estimates for \textit{type of surgical access} ranging between an odds ratio of $3.1$ for RR and of $5.66$ for ML, accompanied by wide confidence intervals but with lower bounds always greater than $1$, see Table \ref{tab:odds}. Given the small events-per-variable ratio of $3.2$ and the small-sample bias-reducing property of FL, the coefficient estimates by FL might be preferable to the ML estimates in this situation although the difference is not substantial. Coefficient estimates by WF, which can be regarded as a compromise between ML and FL, fell between the ones from ML and FL. None of the four adjustment variables showed a significant effect with the methods ML, WF, FL, FLAC and LF, but some of them with CP and RR. This is somewhat in contrast to our simulation results, where confidence intervals from RR were rather more likely to include $0$ than confidence intervals from other methods, independent of the effect size. We explain this discrepancy by one of the assumptions of our simulation study, assuming that in each scenario all explanatory variables had similar effect sizes. Counter to intuition, RR may induce bias away from zero in the effect estimates of irrelevant variables which are correlated to strong predictors.

\begin{table}[htbp]
\caption{Odds ratios with 95\% confidence intervals.} \label{tab:odds}
	\footnotesize
	\centering
	\begin{tabular}{rccccccc}
  \toprule
 & \textbf{ML} & \textbf{WF} & \textbf{FL/FLIC} & \textbf{FLAC} & \textbf{LF} &  \textbf{CP} & \textbf{RR}\\ 
  \midrule
	 \textit{Type of surgical access} & 5.66 & 5.61 & 5.38 & 4.93 & 4.99 & 4.88 & 3.1 \\ 
  (conv. vs. ACD) & (1.89,16.95) & (1.9,17.35) & (1.88,15.97) & (1.73,14.47) & (1.73,14.91) & (2.81,8.48) & (2.07,4.64) \\ 
  \textit{Logistic EuroSCORE } & 1.36 & 1.36 & 1.37 & 1.31 & 1.36 & 1.34 & 1.23 \\ 
  (standardized) & (0.9,2.05) & (0.87,2) & (0.89,1.98) & (0.86,1.9) & (0.86,1.99) & (1.23,1.45) & (1.16,1.31) \\ 
  \textit{Previous cardiac operations} & 3.39 & 3.43 & 3.56 & 2.98 & 2.87 & 2.83 & 2.24 \\ 
  (yes vs.\ no) & (0.79,14.61) & (0.69,13.37) & (0.79,13.02) & (0.64,11.12) & (0.61,10.93) & (1.06,7.54) & (1,5.02) \\ 
	\textit{BMI} & 0.7 & 0.71 & 0.73 & 0.73 & 0.72 & 0.74 & 0.84 \\ 
  (standardized) & (0.39,1.27) & (0.37,1.23) & (0.39,1.25) & (0.41,1.23) & (0.39,1.24) & (0.63,0.86) & (0.77,0.92) \\ 
  \textit{Diabetes} & 1.79 & 1.8 & 1.81 & 1.7 & 1.7 & 1.68 & 1.45 \\ 
  (yes vs.\ no) & (0.57,5.59) & (0.54,5.34) & (0.57,5.18) & (0.53,4.87) & (0.53,4.94) & (0.92,3.06) & (0.93,2.26) \\ 
   \bottomrule
	\end{tabular}
\end{table}

\begin{figure}[htbp]
    \centering
    \includegraphics{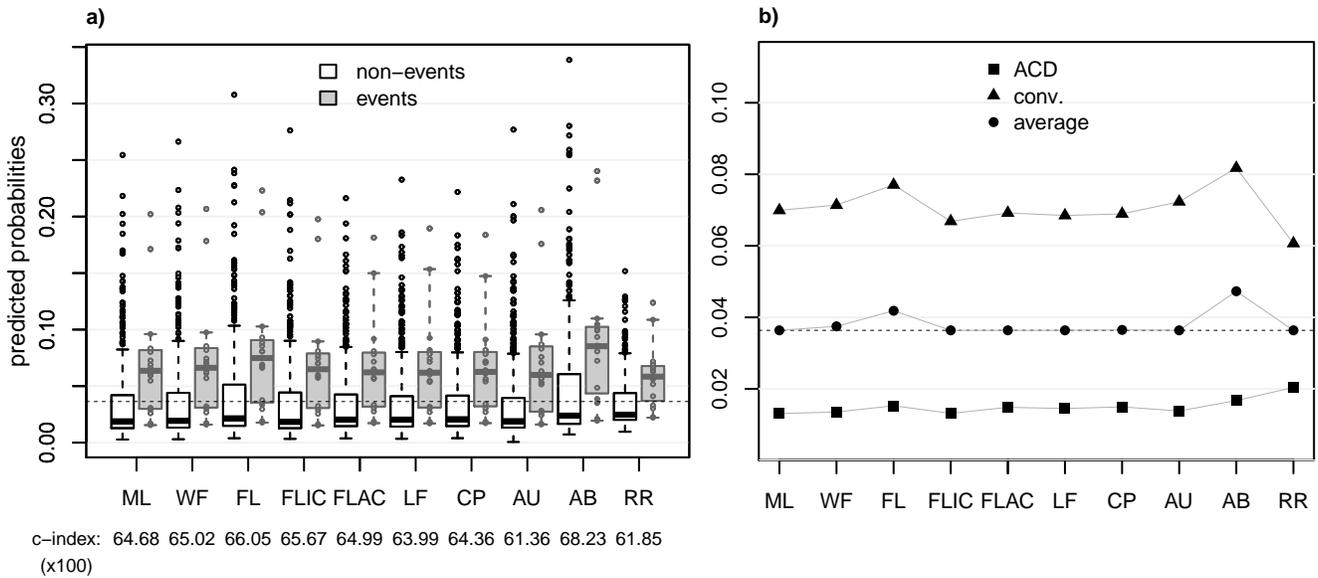}
    \caption{a) Boxplots of predicted probabilities by method and occurrence of event. C-indices were estimated using leave-one-out cross-validation. b) Rectangles and triangles give the predicted probabilities for patients with average covariate values (median logistic EuroSCORE of $5.82$, without previous cardiac operation, median BMI of $26.6$, suffering from diabetes) and with either ACD or conventional surgical access, respectively. Circles mark the average predicted probabilities. In both plots, the horizontal dashed line marks the observed proportion of events (3.6\%).
}\label{fig:boxplotsEvents}
\end{figure}

Figure \ref{fig:boxplotsEvents} a) shows the distribution of the predicted probabilities by method, separately for patients with and without complications. As expected, both FLIC and FLAC pushed the predicted probabilities towards zero, resulting in values clearly smaller than the FL counterparts. On the contrary, AB even increased the FL predictions in magnitude and supplied the largest individual predicted probabilities both, for the event ($24\%$) and non-event group ($33.9\%$). RR had the smallest range of predicted probabilities. The observed proportion of events was overestimated most severely with AB (by $30.1\%$), with FL by $15\%$, with WF by $3.1\%$ and with CP by only $0.3\%$, see Figure \ref{fig:boxplotsEvents} b). The estimated event probabilities for patients with average covariate values (median logistic EuroSCORE of $5.82$, without previous cardiac operation, median BMI of $26.6$, suffering from diabetes) and with either ACD or conventional surgical access shown in Figure \ref{fig:boxplotsEvents} b) exhibit a similar pattern as the boxplots in \ref{fig:boxplotsEvents} a). Again, predicted probabilities by RR had the smallest variability. Predicted probabilities by WF fell between predicted probabilities by ML and FL, whereas for the conventional surgical access group FLIC and FLAC resulted in predicted probabilities smaller than the ones from both, ML and FL. 
Discrimination in terms of cross-validated c-indices (see Figure \ref{fig:boxplotsEvents}) was best for AB with a c-index of $68.2\%$ and poorest for AU with $61.4\%$. \\
All considered methods gave rise to similar conclusions on the role of ACD use, being associated with a significantly lower complication rate than conventional surgical access. For ACD patients with average covariate values, the different methods predicted complication risks between $1.3\%$ and $2\%$, while for comparable conventional access patients predictions ranged between $6.1\%$ and $8.2\%$.

\section{Discussion}

Our simulation study shows that both our suggested methods, Firth-type logistic regression with intercept-correction and Firth-type logistic regression with added covariate, efficiently improve on Firth-type predictions. The complete removal of the bias in the average predicted probability, which amounted up to $19.4\%$ in our simulations, was accompanied by more accurate individual predicted probabilities, as revealed by an RMSE of FLIC and FLAC predicted probabilities ranking at most fourth in $42$ out of $45$ simulation scenarios, only constantly outperformed by RR and never worse than FL. Fortunately, this improvement in predictions has not to be paid for by a lower performance of effect estimation: while for FLIC the effect estimates are identical to those of FL, FLAC introduces some bias but this is compensated by a decreased RMSE. Based on our simulation results, we slightly prefer FLAC over FLIC because of the lower RMSE of predictions. Confidence intervals for coefficient estimates for both, FLIC and FLAC can be easily derived and performed reasonably well in all considered scenarios. Though, in the case of FLIC, the covariance between the intercept and other coefficient estimates can not be based on model output and if required, would have to be estimated by resampling methods. If confidence intervals for predicted probabilities are needed, one could avoid the determination of a full covariance matrix by centering the explanatory variables at the values for which prediction is requested and re-estimating the model. The confidence interval of the intercept can then be interpreted as a confidence interval for the linear predictor. Finally, the two methods can be readily implemented using only some data manipulation steps, ML estimation and an implementation of FL, which is available in software packages such as SAS, R, Stata and Statistica (\cite{logistfSAS}, \cite{firthlogit}, \cite{statistica}).  

The ``weakened'' Firth-type penalization, a compromise between ML and FL estimation, indeed performed better than FL with regard to bias and RMSE of predictions but was outperformed by most of the other methods, by FLIC, FLAC LF, CP and RR. Moreover, reducing the bias in predicted probabilities compared to FL comes at the cost of enlarging the RMSE of effect estimates, as shown by our simulation results in line with intuition. Of course, WF depends essentially on the choice of the weight parameter $\tau$ which was set to $0.1$ as suggested by Elgmati et al., cf.\ \cite{Elgmati15}. Future investigations should clarify whether tuning the weight parameter by optimizing for instance the cross-validated modified $AIC_{\textup{mod}}$, as performed in ridge regression in our study, can make the WF a more attractive option. 

The approximate Bayesian method could not make up for the introduction of bias in predicted probabilities, which is exactly twice as large as in FL estimation, but also performed rather poorly with respect to RMSE. Our results even suggest its inferiority with respect to the approximate unbiased method, giving average predicted probability equal to the proportion of events with often smaller RMSE than AB. Thus, we could not confirm King and Zeng's recommendation to prefer AB over AU ``in the vast majority of applications''. Though, the discrepancy between their and our simulation results advise caution: it seems that none of the two methods is superior to the other in most situations, but that the behaviour strongly depends on the setting. This was also emphasized by a spot-check simulation with balanced outcome, where AB showed lower RMSE than AU (results not shown). One disadvantage of the AU is, that predicted probabilities can fall outside the plausible range of $0$ to $1$. However, the question of deciding between AB and AU might not be a relevant one, since both methods were clearly outperformed by FLIC, FLAC, CP, LF and RR in all simulation scenarios with rare events. It should also be taken into account, that with AB and AU predictions, the analytical relation between linear predictors and predicted probabilities is lost.

The three methods based on weakly-informative priors, penalization by log-$F(1,1)$ priors, by Cauchy priors and ridge regression, do not only differ in the choice of the prior distributions but also in the strategy of data preprocessing. While Greenland and Mansournia, \cite{gm2014}, advocate the use of reasonable, study-independent multiples of SI-units for the LF, Gelman et al., \cite{Gelman08}, suggest to scale continuous variables to have a standard deviation of $0.5$ for CP. In the case of RR, we followed the widespread practice to standardize variables to unit variance and to choose the tuning parameter by the penalized version of the AIC, see for instance \cite{vh1994}. Of course, it is a legitimate question to ask whether other combinations of strategies, for instance penalization by log-$F(1,1)$ priors after stringent scaling of variables or even tuning of prior distribution parameters, might yield better performances, but this would go beyond the scope of this study. Instead, we focused on readily available methods, aimed at routine use in statistical applications. 

Whenever one is willing to accept introduction of bias towards zero for the sake of a small RMSE, RR turned out to be the method of choice, outperforming all other methods with respect to RMSE of coefficients in $38$ and of predictions in $44$ out of $45$ simulation scenarios. However, confidence intervals do not reach their nominal coverage levels because of the combination of bias and reduced variance. The naive approach of deducing Wald confidence intervals from the penalized covariance matrix, which was applied in our simulation study, can not be recommended. In order to avoid these issues in the construction of confidence intervals, CP or FLAC, providing substantially less biased effect estimates than RR with a reasonable RMSE, might be preferable. LF showed a similar performance pattern as CP but was slightly outperformed with regard to the RMSE of coefficients as well as predicted probabilities by CP throughout all simulation scenarios.    

When comparing CP to FLAC, the main difference is the application of independent, univariate priors by CP while a multivariate prior is employed by FLAC. In all methods which use independent priors (including also LF and RR), linear transformations of explanatory variables, such as scaling, or interaction coding, or to achieve orthogonality, will affect predictions from the model. Therefore, \verb+bayesglm+, \cite{arm}, includes an automatic preprocessing of the explanatory variables following \cite{Gelman08}. This preprocessing leads to more stringent penalization of interaction effects than of main effects, which can be desirable in exploratory data analyses. By contrast, in FLAC, any linear transformations of explanatory variables (including different coding) will not affect predictions. Nevertheless, FLAC will apply more shrinkage in larger models, e.g., when interaction effects are included. Thus, with limited data it will penalize complex models more than simple ones. 

Another aspect in this comparison is the interpretation of adjusted regression coefficients. CP, as implemented in \verb+bayesglm+, uses the same prior for the effect of an explanatory variable $X_j$, no matter whether in the model it is adjusted for another variable $X_{k}$ or not. However, the interpretation of the corresponding regression coefficient $\beta_j$ changes fundamentally by adjustment for $\beta_{k}$, in particular if $X_j$ and $X_{k}$ are correlated. Why should then the same prior be imposed, even if only weakly informative, irrespectively of whether $\beta_j$ is adjusted for $\beta_{k}$ or not? In CP, LF or RR (with fixed $\lambda$) the prior definition is not adapted to such changes in interpretation. In FLAC this is implicitly captured by the Jeffreys prior which involves covariances between the coefficients. 

Summarizing, being interested in accurate effect estimates and predictions in the presence of rare events we recommend to use 
\begin{itemize}
\item[-] RR if inference is not required and optimization of the RMSE of coefficients and predicted probabilities is given the priority,
\item[-] FLAC or CP as offering low variability and bias in effect estimates as well as predictions, if valid confidence intervals for effects are needed. We have a slight preference for FLAC because of the invariance properties outlined above. 
\end{itemize}

\section{Appendix}

\begin{table}[!htb]
		\caption{Structure of explanatory variables in the simulation study, following Binder, Sauerbrei und Royston \cite{bs2011}. Square brackets $[ \ldots ]$ indicate that the non-integer part of the argument is removed. $\mathds{1}$ denotes the indicator function, taking value $1$ if its argument is true and $0$ otherwise. } \label{tbl:vars}
			\footnotesize
			\centering
			\begin{tabular}{lllll}
				\toprule
				\textbf{Underlying} & \textbf{Correlation}  & \textbf{Explanatory} & \textbf{Type} & \textbf{Correlation of}\\
				\textbf{variable} & \textbf{of underlying variables} & \textbf{variable} & & \textbf{explanatory variables}\\
				\midrule
				$z_{i1}$ & $z_{i2} \ (0.8), \, z_{i7} \ (0.3)$ &  $x_{i1} = [10z_{i1} + 55]$ & continuous & $x_{i2} \ (-0.6),\, x_{i7} \ (0.2)$\\
				$z_{i2}$ & $z_{i1} \ (0.8)$ & $x_{i2} = \mathds{1}_{\{z_{i2} < 0.6\}}$ & binary & $x_{i1} \ (-0.6)$\\
				$z_{i3}$ & $z_{i4} \ (-0.5), \, z_{i5} \ (-0.3)$ & $x_{i3} = \mathds{1}_{\{z_{i3} \geq -1.2\}}+\mathds{1}_{\{z_{i3} \geq 0.75\}}$ & ordinal & $x_{i4} \ (-0.4), \, x_{i5} \ (-0.2)$\\
				$z_{i4}$ & $z_{i3} \ (-0.5), \, z_{i5} \ (0.5), z_{i7} \ (0.3)$ & $x_{i4} = [\max(0, 100 \exp(z_{i4})-20)]$ & continuous &  $x_{i3} \ (-0.4),\, x_{i5}\ (0.4), \, x_{i7} (0.2),$\\
				& $z_{i8} \ (0.5), \, z_{i9} \ (0.3)$ &&& $x_{i8} \ (0.4), \, x_{i9} \ (-0.2)$\\
				$z_{i5}$ & $z_{i3}\ (-0.3), \, z_{i4} \ (0.5), \, z_{i8}\ (0.3),$ & $x_{i5} = [\max(0, 80 \exp(z_{i5})-20)]$ & continuous & $x_{i3}\  (-0.2), \, x_{i4} \ (0.4), \, x_{i8}\ (0.2),$\\
				& $z_{i9} \ (0.3)$ &&& $x_{i9}\ (-0.2)$\\
				$z_{i6}$ & $z_{i7} \ (-0.3),\,  z_{i8} \ (0.3) $ &$x_{i2} = \mathds{1}_{\{z_{i6} < -0.35\}}$ & binary & $x_{i7} \ (0.2), \, x_{i8} \ (-0.2)$\\
				$z_{i7}$ & $z_{i1} \ (0.3), \, z_{i4} \ (0.3), \, z_{i6} \ (-0.3)$ &$x_{i7} = \mathds{1}_{\{z_{i3} \geq 0.5\}}+\mathds{1}_{\{z_{i3} \geq 1.5\}}$ & ordinal & $x_{i1} \ (0.2),  \, x_{i4} \ (0.2), \, x_{i6} \ (0.2)$\\
				$z_{i8}$ & $z_{i4} \ (0.5), \, z_{i5}\ (0.3),\, z_{i6} \ (0.3)$  &$x_{i8} = [10z_{i8} + 55]$ & continuous & $x_{i4} \ (0.4), \, x_{i5}\  (0.2), \, x_{i6}\ (-0.2),$\\
				& $z_{i9} \ (0.5)$ &&& $x_{i9}\ (-0.4)$\\
				$z_{i9}$ & $ z_{i4} \ (0.3),\, z_{i5} \ (0.3),\, z_{i8} (0.5)$  & $x_{i9} = \mathds{1}_{\{z_{i9} < 0\}}$ & binary & $x_{i4}\ (-0.2), \,x_{i5}\ (-0.2), \, x_{i8}\ (-0.4)$\\
				$z_{i10}$ & $\;-$ & $x_{i10} = \mathds{1}_{\{z_{i10} < 0\}}$ & binary & $\;-$ \\
				\bottomrule
			\end{tabular}
\end{table}

\noindent \textbf{Acknowledgements:} We thank Michael Schemper and Mohammad Ali Mansournia for helpful comments. The data on complications in cardiac surgery were kindly provided by Paulo A. Amorim, Alexandros Moschovas, Gloria F\"arber, Mahmoud Diab, Tobias B\"unger and Torsten Doenst from the Department of Cardiotheracic Survergy at the University Hospital Jena. The research leading to these results has received funding from the Austrian Science Fund (FWF) within project I 2276 and from the Slovenian Research Agency (ARRS) within project N1-0035.

\bibliographystyle{wileyj}

\bibliography{LogReg}

\end{document}